\documentclass[12pt]{article}

\usepackage{graphicx}
\usepackage{amssymb}
\usepackage{amsmath}
\usepackage{amsfonts}
\newcommand{\be}{\begin{equation}}
\newcommand{\ee}{\end{equation}}
\newcommand{\bea}{\begin{eqnarray}}
\newcommand{\eea}{\end{eqnarray}}
\newcommand{\nn}{\nonumber}
\newcommand{\cO}{{\mathcal{O}}}
\newcommand{\cT}{{\mathcal{T}}}
\newcommand{\ri}{{\rm i}}
\begin{document}

\title{Anharmonic oscillator: A playground to get insight into renormalization}

\author{Saman Moghimi-Araghi\thanks{E-mail address: samanimi@sharif.edu }~ and
Farhang Loran\thanks{E-mail address:
loran@iut.ac.ir}\\[6pt]
$^{*}$Department of Physics, Sharif University of Technology, \\ Tehran, P.O. Box:11555-9161, Iran \\[6pt]
$^\dagger$Department of Physics, Isfahan University of Technology, \\ Isfahan 84156-83111, Iran}

\date{ }
\maketitle

\begin{abstract}
In the presence of interactions the frequency of a simple harmonic oscillator deviates from the noninteracting  one. Various methods can be used to compute the changes to the frequency  perturbatively. Some of them  resemble the methods used in quantum field theory where the bare values of the parameters of the theory run when an interaction is added. In this paper, we review some of these techniques and introduce some new ones in line with quantum field theory methods. Moreover, we investigate the case of more than one oscillator to see how the frequencies of small oscillations change when non-linear terms are added to a linear system and obsereve an interesting beat phenomenon in degenerate coupled oscillatory systems.

\vspace{6pt}





\end{abstract}
\tableofcontents

\section{Introduction}
The harmonic oscillator plays a central role in various areas of physics. The description of many phenomena can be reduced to the problem of a simple harmonic oscillator (SHO) or a collection of such oscillators. These phenomena range from optics and electromagnetic waves to the vibrations of a solid crystal and the quantum field theory (QFT). However, in most cases, the SHO  is only an approximation that helps us  solve the problem analytically and discover  the main features of the model. Therefore, many techniques have been developed to study the perturbation of this simple periodic motion. Perhaps the simplest nonlinear variation of SHO is the Duffing equation, where only a cubic term is added to the original SHO equation of motion \cite{DuffingPaper}. The Duffing equation has been thoroughly studied analytically, perturbatively, and numerically in many different variations  including driving force and damping terms \cite{DuffingBook,NayfehBook}.

One of the most interesting features of these perturbed systems is the dependence of the physical quantity ``period of oscillation'' on the amplitude of the motion, \cite{FulcherandDavis(1976),GilandGregorio(2006),Zilio(1982)}. Different approaches have been devised to compute this dependence \cite{FulcherandDavis(1976),GilandGregorio(2006),StrogatzBook,BenderBook,Bender,VerhulstBook, KahnandZarmi(2000),GilmartinKleinandLi(1979),Amore1,Duki,Fernandez,O'Shea,Roberts}.  Two of the best known methods are the  ``multiple scale analysis''  and the Poincar\'{e}-Lindstedt method. Both methods rely on removing the secular terms that appear in the perturbation expansion. In the method of multiple scales, as the name implies, the motion is divided into different parts, each of which has a different time scale; one part has the time scale of the SHO, and  there is another part that varies much slower, and so on \cite{BenderBook,Jakobsen,Sharp}. The requirement of removing the secular term helps to find various time scales, which culminate to the  change of the period of oscillation.  In the Poincar\'{e}-Lindstedt method, the changes in frequency are taken into account from the beginning and they are computed order by order perturbatively \cite{VerhulstBook,Rand,Amore}.

Some of these methods have  similarities to the renormalization method in QFT \cite{PeskinBook,Weinberg,Berestetskii,Sterman}. Usually when an interacting field theory is treated perturbatively, one encounters some infinities in the subleading terms. These infinities are removed by letting some of the physical quantities, such as the mass of the particles and the coupling constants, ``run'' with the scale of energy. Student who are learning the renormalization procedure in QFT for the first time, may not understand why we have to change these physical quantities when an interaction is switched on.   In general, renormalization of a QFT requires lengthy calculations which might conceal its physical implications. In this paper, we apply similar methods   to the simple problem of SHO to see how they  work in  a familiar environment without getting involved in too many complications, hoping that this gives an insight to the same phenomenon in QFT.

In the following, we first review the anharmonic oscillator through the example of the motion of a pendulum, then review the previously known approaches to finding the changes of the period of oscillations and introduce a few other approaches that resemble  those usually used in QFT. Finally, we consider the case of coupled anharmonic  oscillators where we find some interesting phenomena in the degenerate cases.

\section{Anharmonic oscillator: one degree of freedom  from various vantage points }

Consider the equation of motion of a  simple pendulum which is written in the following form:

\be
    \label{pendulum}
    \ddot\theta+\frac{\mathfrak{g}}{L}\sin\theta=0,
    \ee
where $L$ is the pendulum length, $\mathfrak{g}$ denotes the gravitational acceleration, and theta indicates the angle that the cord makes with the vertical. Here we are using the ``dot notation'' and denote  the time derivative of a dependent variable by placing a dot over it, so $\ddot\theta:= d^2\theta/dt^2$ stands for the angular acceleration. Oscillations with small amplitudes are ruled by the equation of an SHO
    \begin{align}
    \label{pendulum-first}
    &\ddot\theta+\omega_0^2\theta=0,&\omega_0:=\sqrt{\frac{\mathfrak{g}}{L}}.
    \end{align}
Thus, the frequency of such small oscillations is $\omega_0$. To obtain this equation we have used the Taylor series
    \be
    \label{sin}
    \sin\theta=\theta-\frac{\theta^3}{6}+\cdots,
    \ee
and considered only the linear term. This approximation is plausible only if the oscillation amplitude is small.  On the contrary, if we let the pendulum to start its motion from an initial angle which is not small, equation \eqref{pendulum-first} would not be a good approximation.

It is known that all possible solutions to the Eq.\eqref{pendulum} are oscillatory, though the oscillations might not be harmonic.\footnote{Physically we can suppose that the oscillation amplitude is smaller than $\frac{\pi}{2}$ to guarantee that in a real setup the pendulum's wire remains tight during the motion and consequently equation \eqref{pendulum} is valid.} In fact Eq.\eqref{pendulum} is integrable whose solution can be given in terms of elliptic integrals. We are not going to discuss the exact solution here. Instead, we focus on the frequency of the oscillation and  figure out its dependence on the amplitude. We try various perturbation methods to obtain the solution. However, as we are mainly interested in the technicalities of these methods and their similarities to the methods of  QFT, we only keep the subleading term in the Taylor series of  $\sin\theta$ about $\theta=0$ and suppose that the equation of motion is ``exactly''
    \be
    \label{main}
    \ddot x+\omega_0^2 x+\lambda x^3=0,
    \ee
where we have replaced $\theta$ by $x$. Eq.\eqref{main} is known as the undamped Duffing equation. Although for the pendulum we can determine $\lambda$ in terms of $\omega_0$ by means of the Taylor series, but in the following we consider it as an independent parameter.

Before moving on, let us write the field equation for the ``$\lambda \varphi^4$'' theory in QFT to manifestly see the similarities of the two theories:
\be
\ddot\varphi-c^2\nabla^2\varphi+\omega_0^2\varphi+\lambda\varphi^3=0,
\ee
where $c$  denotes the speed of light, $\hbar \omega_0/c^2$ is interpreted as the ``bare'' mass of the scalar field $\varphi$, $\hbar$ is  the Planck constant, and $\lambda$ is the coupling constant. Thus, the time-dependence of those scalar fields which solve the Laplace equation $\nabla^2\varphi=0$ is similar to $x$ solving Eq.\eqref{main}.

We are going to treat   $\lambda$ as the perturbation parameter, although it is not necessarily  dimensionless. In fact the only parameters in the equation of motion are $\omega_0$ and $\lambda$ and it is not always possible to make a dimensionless combination out of them.  To clarify what the smallness of $\lambda$ actually means, we need to take into account the constant of motion, or the total ``energy:''
    \be
    E:=\frac{{\dot x}^2}{2}+\frac{\omega_0^2x^2}{2}+\frac{\lambda x^4}{4},
    \ee
where we have set the mass of the bob equal to unity ($m=1$).

Now we turn back to the pendulum system. The amplitude of oscillations, $R$, can be obtained from the mechanical energy by  solving the following equation\footnote{
 If  $\lambda<0$,  some of the initial conditions with large amplitudes  do not result in oscillation. Therefore, we consider oscillations whose amplitudes are small enough so that the contribution from the $x^4$ term is negligible and can be treated perturbatiely.}
    \be
    \label{energy}
    E=\frac{\omega_0^2R^2}{2}+\frac{\lambda R^4}{4},
    \ee
which gives
    \begin{align}
    &R=\sqrt{\frac{E}{\omega_0^2}}\sqrt{\frac{1}{1+\sqrt{1+u}}}\,,& u:=\frac{4\lambda E}{\omega_0^2}.
    \end{align}
$u$ is a dimensionless quantity linearly dependent on $\lambda$. Thus, we consider it as the perturbation parameter and require that  $\left|u\right|\ll1$. Therefore, the precise meaning of the perturbation in $\lambda$ is the perturbation in $u$.

In addition to $u$ we can define other dimensionless combinations of the parameters of the problem. The dimensions of $E$ and $\omega_0^2R^2$ are the same. In fact we are studying the problem in the regime $E\approx\omega_0^2R^2$. Therefore, instead of $u$ we could use $\tilde{u}:=uf\left(\frac{\omega_0^2R^2}{4E}\right)$ for an arbitrary function $f$. In particular $f(x)=x^2$ gives $\tilde{u}=\frac{\lambda R^4}{4E}$ which we encounter frequently in the course of this study.

In the following, we try to solve Eq.\eqref{main} perturbatively and see how simple perturbation fails and why we need to renormalize the frequency of the oscillations when the total energy is not small enough.  We present a few different ways to compute $\omega$,  the frequency of oscillatory solutions of \eqref{main}, for small values $E$ by perturbation. We already know that the leading term is $\omega_0$. In the following we  compute $\delta\omega:=\omega-\omega_0$ in terms of the coupling $\lambda$ and the amplitude $R$. Instead of trying to compute  higher order terms,  we  mainly focus on the leading term and save our effort to explore different approaches to the perturbation, especially those which have connections with QFT.

\begin{figure}[t]
\centering
  \includegraphics[width=0.5\linewidth]{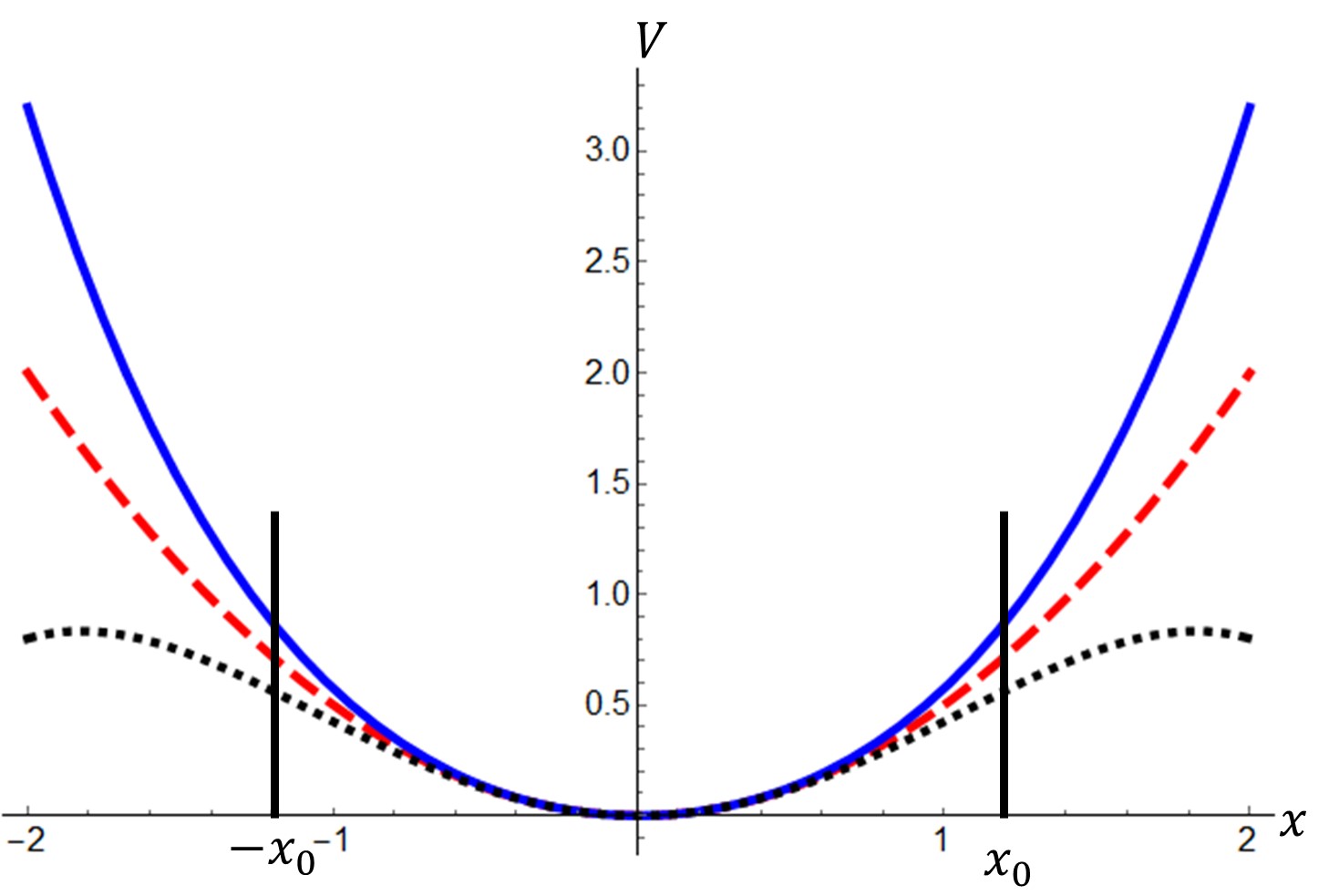}\\
  \caption{The potential $V(x)= x^2/2 + \lambda x^4/4$ for $\lambda=+0.3$ (solid, blue),
$\lambda=0$ (dashed, red) and $\lambda=-0.3$ (dotted, black). The oscillations with the
three potentials but with the same amplitude $x_0$ is considered. The system with
$\lambda=+0.3$ has the greatest energy and its mean value of the kinetic energy is the largest during the
oscillations and hence has the shortest period.}\label{Potential}
\end{figure}

To have a feeling about how this frequency depends on the amplitude, let us have a look at Figure \ref{Potential} where we have depicted $V(x)$ for $\omega_0=1$ and $\lambda=0,\pm0.3$. For low energies the difference in $V(x)$ and hence in $x(t)$ would be negligible. For higher values of energy,  the change in the  $x(t)$ would be manifest. Let us fix the amplitude of  oscillations and suppose that it equals $x_0$ as shown in the figure. It is clear that the mechanical energy for $\lambda=0.3$  is bigger  that the mechanical energy for $\lambda=0$ which is larger than the mechanical energy for $\lambda=-0.3$. Thus, in the region where these three potentials are more or less equal, the average velocity of the particle moving on the $\lambda=0.3$ curve is higher than the average velocity of  the particle moving on the $\lambda=-0.3$ curve. Therefore, for small values of the mechanical energy the  frequency should be an  increasing function of $\lambda$. In the following we give a quantitative account of this phenomenon.

\subsection{Simple Integration}\label{section-simple-integration}
Similarly to all classical systems with one degree of freedom in which the mechanical energy is conserved, equation \eqref{main} presents an integrable system. Eq.\eqref{energy} can be solved for the velocity
    \be
    {\dot x}^2=\frac{\lambda}{2}(R^2-x^2)\left(x^2+\frac{4E}{\lambda R^2}\right).
    \ee
The existence of the roots $x=\pm R$ is quite natural since there are two turning points symmetrically located on the $x$ axis where, by definition,  the velocity  is zero. This equation gives the velocity as a function of $x$. This  equation can be used to compute the time elapsed in moving from one turning point to the other. Therefore, the period is given by
    \be
    T(\lambda)=2\int_{-R}^R \frac{dx}{\left|\dot x\right|}=4\int_0^R\frac{dx}{\sqrt{\frac{\lambda}{2}(R^2-x^2)\left(x^2+\frac{4E}{\lambda R^2}\right)}}.
    \ee
This is an elliptic integral as can be realized by introducing a new variable $\phi:=\frac{x}{R}$
    \begin{align}
    \label{T}
    &T(\lambda)=\frac{4R}{\sqrt{2E}}\int_0^{\frac{\pi}{2}}\frac{d\phi}{\sqrt{1+\alpha\cos^2\phi}}\,,&\alpha:=\frac{\lambda R^4}{4E}.
    \end{align}
As we have discussed before, $\alpha$ is a plausible  perturbation parameter. Thus,  we can use the Taylor series of the integrand on the right hand side of \eqref{T} to obtain the corresponding Taylor series of $T(\lambda)$:
    \be
    \label{T-sol}
    T(\lambda)=\frac{4R}{\sqrt{2E}}\int_0^{\frac{\pi}{2}}d\phi\left(1-\frac{\alpha}{2}\cos^2\phi+\cO(\alpha^2)\right)
    =\frac{2\pi R}{\sqrt{2E}}\left(1-\frac{\alpha}{2}+\cO(\alpha^2)\right).
    \ee
 To obtain the frequency $\omega=2\pi T^{-1}$ we only  need to use \eqref{energy} to compute  the right hand side of \eqref{T-sol} and keep the linear term in $\lambda$ in the corresponding Taylor series:
    \be
    \label{omega2}
    \omega^2=\omega_0^2\left(1+\frac{3}{4}\frac{\lambda R^2}{\omega_0^2}+\cO(\lambda^2)\right).
    \ee
This result clearly shows that at low energies (small amplitudes) $\omega$ is an increasing function of $\lambda$.

The higher order terms can be also computed by this method. For example, the next to the leading order term is given by
    \be
    \label{second-order}
    \omega^2=\omega_0^2\left[1+\frac{3}{4}\frac{\lambda R^2}{\omega_0^2}-\frac{3}{128}\left(\frac{\lambda R^2}{\omega_0^2}\right)^2+\cO(\lambda^3)\right].
    \ee
There are a few theqniques in the literature to improve the convergence of such perturbative expansions \cite{Duki,Fernandez}. Although this solution is more accurate than \eqref{omega2} for the problem defined by \eqref{main}, the $\lambda^2$ term can not be trusted for  the motion of the pendulum, because Eq.\eqref{main} only captures the next to leading order term in the series \eqref{sin}. Furthermore \eqref{sin} implies that for the pendulum  $\lambda=-\omega_0^2/6$ and consequently
    \be
    \omega^2=\omega_0^2\left(1-\frac{R^2}{8}+\cO(R^4)\right).
    \ee
Assuming that $\mathfrak{g}=9.8\,{\rm m}/{\rm s}^2$ precisely, for  a pendulum whose length is exactly $L=0.5\,{\rm m}$ we  obtain $T=1.41923\,{\rm s}$ for small oscillations, while for $\theta_0= 30^\circ$ Eq.\eqref{omega2} gives  $T=1.44397\,{\rm s}$ which is about $2$ percent larger. The precise value of the period for $\theta_0= 30^\circ$ given by  Eq.\eqref{T-sol} is $T\simeq 1.44393\,{\rm s}$ implying that the relative error of  the first order calculation is $\sim10^{-5}$.
\subsection{Naive perturbation: Rise of non-periodic terms}\label{section-Naive}
When we are planning for calculating the small deviations in the trajectories of a physical system as a consequence of small changes in the equations of motion, the perturbation method is the first thing that rings a bell. Assume that we have the solution $x_0(t)$ to the equation $\mathcal{L}_0(x(t))=0$ but we are supposed to solve another equation
    \be
    \label{L}
    \mathcal{L}_0(x)+\lambda \mathcal{W}(x)=0,
    \ee
where $0<\lambda\ll1$. Since $x_0(t)$ solves $\mathcal{L}_0(x)=0$,  we use  the anstaz
    \be
    \label{ansatz}
    x(t)=x_0(t)+\lambda\, x_1(t)+\lambda^2x_2(t)+\cO(\lambda^3),
    \ee
in  \eqref{L} and use the Taylor series to  rewrite $\mathcal{L}_0(x)+\lambda \mathcal{W}(x)$ as a polynomial in $\lambda$ whose  zeroth order term is zero because $x_0(t)$ solves $\mathcal{L}_0(x)=0$. Therefore, Eq.\eqref{L} becomes a polynomial equation. The lowest order term  gives $x_1$ in terms of $x_0$ and the $\lambda^{i+1}$ term gives $x_{i+1}$ in terms of $x_j$, $j<i$.

For equation \eqref{main} $\mathcal{L}_0(x)=\ddot x+\omega_0^2x$, $\mathcal{W}(x)=x^3$ and
    \be
    \label{x0}
    x_0=A\cos(\omega_0t+\phi).
    \ee
Using the ansatz \eqref{ansatz} in \eqref{main} we obtain
    \be
    \label{x1}
    \ddot x_1+\omega_0^2x_1=-x_0^3.
    \ee
We only need the particular solution of this equation because the constants  $A$ and $\phi$ are enough to compensate for the initial values of $x$ and $\dot x$.

Before solving $x_1$ let us recall our intention which is  a perturbative calculation of the deviation in the frequency of oscillation from $\omega_0$ as a consequence of the perturbation generated by the $\lambda x^4$ term. Therefore, we are seeking an oscillatory solution, but is the solution to \eqref{x1} periodic? To find the answer, we note that
    \be
    \label{cos}
    \cos^3(\omega_0t+\phi)=\frac{3\cos(\omega_0t+\phi)+\cos[3(\omega_0t+\phi)]}{4}.
    \ee
So the external force term on the right hand side of \eqref{x1} is a combination of two terms and consequently $x_1$ is a combination of the responses to these two stimuli too. However, the particular solution to the equation
    \be
    {\ddot x}_1+\omega_0^2x_1=-\frac{3A^3}{4}\cos(\omega_0t+\phi),
    \ee
is not periodic; it is proportional to $t\sin(\omega_0t+\phi)$  which is known as the secular term. This is an oscillation whose amplitude grows linearly with time. We already now that $x(t)$ is periodic and we are going to compute its frequency. We would like to draw the reader's attention  to the point that some similar problems arise in QFT when we keep the bare parameters of the  system unaltered.  In the following, we review two celebrated approaches to dismiss the secular term \cite{NayfehBook,BenderBook}.
\subsection{Multiple-scale analysis}
We have seen that a naive approach to perturbation can be misleading. A successful method in analyzing dynamical systems is the multiple scale analysis. In this approach, which is often applied to oscillating systems, we assume that there are two time scales in the theory. One of them is the natural time scale given by $\omega_0$ and the other one $\cT_0\gg\omega_0^{-1}$ is introduced by the perturbation. In particular, in the case of anharmonic oscillator, we shall express the motion in the following way: the particle is oscillating with frequency $\omega_0$ while the amplitude is oscillating with  frequency $\sim \cT_0^{-1}$. To separate the evolution with respect to these two time scales, it is assumed that there exist two independent time variables
    \begin{align}
    \label{tau-cT}
    &\tau:=t,&\cT:=\lambda\, t,
    \end{align}
Using the anstaz
    \be
    x(t;\lambda)=x_0(\tau,\cT)+\lambda x_1(\tau,\cT)+\cdots,
    \ee
together with
    \be
    \frac{d}{dt}=\partial_\tau+\lambda\,\partial_\cT,
    \ee
where
    \begin{align}
    &\partial_\tau:=\frac{\partial}{\partial\tau}, &\partial_\cT:=\frac{\partial}{\partial\cT},
    \end{align}
in \eqref{main} and keeping only linear  terms in  $\lambda$ in the equation of motion, we obtain
    \be
    \left(\partial_\tau^2x_0+\omega_0^2 x_0\right)+\lambda\left(\partial_\tau^2x_1+\omega_0^2 x_1+2\partial_\tau\partial_\cT x_0+x_0^3\right)+\cO(\lambda^2)=0.
    \ee
The zeroth order term  gives
    \be
    \label{x0-BC}
    x_0(\tau,\cT)=A(\cT)\cos\left(\omega_0\tau+\phi(\cT)\right)=B(\cT)\cos\omega_0\tau+C(\cT)\sin\omega_0\tau.
    \ee
Note that we have allowed $A$ and $\phi$ to depend on $\cT$ because the equation of motion for $x_0(\tau,\cT)$  determines only its $\tau$ dependence and leaves its $\cT$ dependence arbitrary. Consequently
    \begin{align}
    &B:=A\cos\phi,&C:=-A\sin\phi,
    \end{align}
are undetermined functions of $\cT$. To proceed, we write the equation of motion for $x_1$ as
    \bea
    \partial_\tau^2+\omega_0^2 x_1&=&-2\partial_\tau\partial_\cT x_0-x_0^3\\
    &=&-2\omega_0\left[-\partial_\cT B\sin\omega_0\tau+\partial_\cT C\cos\omega_0\tau\right]-\left(B\cos\omega_0\tau+C\sin\omega_0\tau\right)^3.
    \eea
Some of the terms appearing on the right hand side of the above equation oscillate with frequency $\omega_0$ and we know that they would result in a secular contribution to $x_1$. To obtain an oscillatory solution, we need to choose $B(\cT)$ and $C(\cT)$ in such a way that such terms cancel out. These terms appear in the following combination
    \be
    \left[2\omega_0\partial_\cT B-\frac{3C}{4}(B^2+C^2)\right]\sin\omega_0\tau-\left[2\omega_0\partial_\cT C+\frac{3B}{4}(B^2+C^2)\right]\cos\omega_0\tau,
    \ee
which is zero only if
    \begin{align}
    \label{BC}
    &\partial_\cT B=\frac{3C}{8\omega_0}(B^2+C^2),&\partial_\cT C=-\frac{3B}{8\omega_0}(B^2+C^2).
    \end{align}
Thus, we have  $B\partial_\cT B+C\partial_\cT C=0$  implying that $R:=\sqrt{B^2+C^2}$ is constant. Using this result in \eqref{BC} we obtain
    \begin{align}
    &B=R\cos(\nu\cT+\phi),&C=-R\sin(\nu\cT+\phi),
    \end{align}
where
    \be
    \label{nu}
    \nu:=\frac{3R^2}{8\omega_0^2}.
    \ee
Collecting everything  and using trigonometric equalities, we verify that
    \be
    \label{x0-2time}
    x_0=R\cos(\omega_0\tau+\nu\cT+\phi)=R\cos(\kappa\omega_0t+\phi)
    \ee
in which
    \be
    \label{kappa}
    \kappa:=1+\frac{3\lambda R^2}{8\omega_0^2}.
    \ee
This is exactly what we have obtained in section \eqref{section-simple-integration}, cf. \eqref{omega2}. Introduction of two time variables has applications in various problems, but its extension to compute higher order terms is complicated because that requires more time variables.

\subsection{Poincar\'{e}-Lindstedt method: It looks like renormalization}\label{section-PL}
The method that we study in this section is a standard approach to perturbation. At  its roots it is  similar to  our approach  in section \ref{section-Naive} which led us to the  secular term. In the previous section, we removed the secular term by introducing two time scales one for the oscillation of the particle and the other one for the changes in the amplitude. We found out that the combination of these two effects is a modification in $\omega_0$. However, in the Poincar\'{e}-Lindstedt  method  we let the frequency of oscillations deviate from $\omega_0$ from the beginning. Actually, the harmonic oscillator (when $\lambda=0$) has frequency of oscillation equal to $\omega_0$ which  is usually called the bare parameter. This bare parameter is just a parameter that defines the theory, however, the physical properties of the system, such as the frequency of oscillations, might be different from the bare parameter when the nonlinear term is added to the system. Therefore, to find a proper perturbative solution one should manage to compute this physical quantity too.

This way of thinking  has been successfully pursued  in various problems in physics, especially in QFT.  In general, we begin with a set of linear field equations representing free (i.e., non-interacting) fields. These equations include  some parameters such as the rest mass of the corresponding particle. So this is a parameter that defines the theory. When interactions are added to the theory by means of some other parameters known as the coupling constants, the physical properties of the field, like the rest mass, might become different from the bare parameter that defines the theory. In this picture, the interacting fields are  considered as perturbations to ``freely oscillating'' states and  the perturbations result in deviation of the physical properties from the initial parameters of the theory. Historically, these changes were discovered as a consequence of the regularization. Take the standard model of particle physics as example. When we compute  the first order terms in the perturbation series, some of the results become infinite. The procedure of removing these infinities is called regularization, whose leftover is a change in the bare parameters of the theory. Actually, these infinities appear if we do not recognize that the actual mass could be different from the mass parameter of the noninteracting theory.  Fortunately, in the simple problem of anharmonic oscillator, we do not encounter infinities; we only need to remove the secular term.

To start, we rewrite equation \eqref{main} as follows
    \be
    \label{main-1}
    \ddot x+\omega^2x-(\omega^2-\omega_0^2)x+\lambda x^3=0.
    \ee
That is we  suppose that the ``correct'' theory has a frequency $\omega$ different from the bare value  $\omega_0$. In QFT, the term $(\omega^2-\omega_0^2) x$ is called the counter-term. It is clear that for $\lambda=0$ Eq. \eqref{main-1} has only one oscillating  solution $x_0$ whose frequency is $\omega_0$. So any deviation from $\omega_0$ is a consequence of the presence of the nonlinear term $\lambda x^3$ implying that
    \be
    \label{delta-omega}
    \omega^2-\omega_0^2=\lambda\mu+\cO(\lambda^2).
    \ee
We should determine $\mu$ in such a way that the solution of $x$ remains oscillating after adding the first order correction. By using \eqref{ansatz} and \eqref{delta-omega} in \eqref{main-1} and rewriting the result as a polynomial equation in $\lambda$ we obtain
    \begin{align}
    &\label{x0-PL}
    \ddot x_0+\omega^2 x_0=0,\\
    &\label{x1-PL}
    \ddot x_1+\omega^2x_1=\mu x_0-x_0^3.
    \end{align}
The solution to \eqref{x0-PL} is
    \be
    x_0=A\cos(\omega t+\phi).
    \ee
Using this result  in \eqref{x1-PL} we obtain
    \be
    \ddot x_1+\omega^2x_1=\left(\mu -\frac{3A^2}{4}\right)A\cos(\omega t+\phi)-\frac{A^3}{4}\cos[3(\omega t+\phi)],
    \ee
where we have used the trigonometric identity \eqref{cos}. Choosing
    \be
    \mu =\frac{3A^2}{4},
    \ee
the source of the secular term vanishes. This choice reproduces \eqref{kappa} and we have
    \be
    \ddot x_1+\omega^2x_1=-\frac{A^3}{4}\cos[3(\omega t+\phi)],
    \ee
whose particular solution is
    \be
    x_1=\frac{A^3}{32\omega^2}\cos[3(\omega t+\phi)].
    \ee
In this way, we  determine how  the nonlinear term $\lambda x^3$ changes the frequency of the oscillation to $\omega$ and how it deviates from the bare frequency $\omega_0$. Moreover, we observe that the motion is not a simple oscillation anymore and a new term whose frequency is $3\omega$ has emerged.

This method can be more easily used to compute the higher order terms contrary to the multiple scale method. We just need to keep the $\lambda^2$ corrections both in $\omega^2-\omega_0^2$ and in $x(t)$ and determine the former by allowing only  oscillatory contributions to the latter.
\subsubsection{A rule of thumb}\label{thumb}
Motivated by the Poincar\'{e}-Lindstedt method, we memorize the following algorithm for solving \eqref{main}. We try the anstaz
    \be
    x=A\cos(\omega t +\phi)+\lambda B\cos({\tilde{\omega}} t +\theta)+\cO(\lambda^2),
    \ee
where
    \begin{align}
    & \lim_{\lambda\to 0}\omega= \omega_0,&\lim_{\lambda\to 0}{\tilde{\omega}}\neq \omega_0.
    \end{align}
Using this in \eqref{main} and dropping terms nonlinear in $\lambda$  we obtain
    \be
    (\omega_0^2-\omega^2)A\cos(\omega t +\phi)+\lambda(\omega_0^2-{\tilde{\omega}}^2)B\cos({\tilde{\omega}} t +\theta)=-\frac{\lambda A^3}{4}\Big(3\cos(\omega t +\phi)+\cos(3\omega t +3\phi)\Big),
    \ee
which determines $B$, $\theta$, ${\tilde{\omega}}$ in terms of $A$, $\phi$ and $\omega_0$, and   especially gives
    \be
    \omega^2-\omega_0^2=\frac{3\lambda A^2}{4}.
    \ee

\subsection{Green function and the Fourier space}

In the previous section, we noted that the method to overcome the secular term resembles  the one used in field theory to dismiss the infinities.  One of the  common tools in field theory is the Green function. For example, the poles of this function determine the mass of the particle. Interactions change the position of the poles. Thus the  physical mass becomes different from the bare mass. In this section, we want to devise a similar method for the anharmonic oscillator.

In studying an oscillating system, it is quite natural to work in the Fourier space. In general, the Fourier transforms of  linear differential equations are  linear algebraic equations whose solutions can be obtained easily. Take the SHO  $\ddot x+\omega_0^2x=0$ for instance. To find the Green function, an impulse at $t=0$ can be added to the equation of motion   according to
    \be
    \label{impulse}
    \ddot x+\omega_0^2x-v_0\,\delta(t)=0,
    \ee
where $\delta$ denotes the Dirac delta function. The $\delta$-function implies that $x$ oscillates freely for $t\neq0$ but it is driven by a force $f(t)$ during a short period $t\in(-\epsilon,\epsilon)$ for $\epsilon\to 0$ such that the resulting change in the momentum is finite.
    \be
    \lim_{\epsilon\to0}\int_{-\epsilon}^\epsilon dt\,f(t)=mv_0.
    \ee
$m$ is the mass of the particle which we have assumed to be unity throughout this paper. The limit  $\epsilon\to 0$ practically means that the duration of the impetus is negligible in comparison with the period of oscillation. For a particle initially at rest, the solution to the equation of motion is easy:
    \be
    x(t)=\left\{\begin{array}{ll}0& t<0,\\\frac{v_0}{\omega_0}\sin(\omega_0 t)&t>0.
    \end{array}\right.
    \ee
To rewrite this Green function in the Fourier space, we can start with computing the Fourier transform of $x(t)$ which is given
by
    \be
   x(t)\to \tilde{x}(\Omega):=\int_{-\infty}^\infty dt\, e^{-\ri\Omega t}x(t),
    \ee
whose inverse is
    \be
    \label{Fourier-inv}
    x(t)=\int_{-\infty}^\infty\frac{d\Omega}{2\pi}e^{\ri\Omega t}\tilde{x}(\Omega).
    \ee
Thus the Fourier transform of \eqref{impulse} reads
    \be
    (-\Omega^2+\omega_0^2)\tilde{x}(\Omega)-v_0=0,
    \ee
whose solution is
    \be
    \label{delta-1}
    \tilde{x}(\Omega)=\frac{v_0}{\omega_0^2-\Omega^2}.
    \ee
We see that the poles of this Green function in Fourier space show the frequency of oscillation which for the SHO  equals  $\omega_0$.

We can use this Green function and make an inverse Fourier transform to obtain the position of the oscillator
    \be
    \label{contourEQ}
    x(t)=-\frac{v_0}{2\omega_0}\int_{-\infty}^\infty\frac{d\Omega}{2\pi}e^{i\Omega t}\left(\frac{1}{\Omega-\omega_0}-\frac{1}{\Omega+\omega_0}\right).
    \ee
This integral is not well-defined since the real line  passes through the singularities at $\Omega=\pm\omega_0$. In the vicinity of each pole we need to detour slightly upward or downward in the complex $\Omega$ plane to avoid the poles. The detour should be compatible with causality. Recall that the particle is at rest for $t<0$ and its oscillation at $t>0$ is caused by an impulse at $t=0$. Therefore, for $t<0$ the causal relation between the impulse and the oscillation requires that in the vicinity of both poles we take the counterclockwise roundabout route because if we complete the contour by taking a clockwise tour at infinity in the lower half plane, we end up with a  closed integration contour that does not enclose any singularity.  Therefore, the Cauchy's residue theorem implies that the integral around the whole of the contour is zero. Since the integral along  the clockwise piece at infinity is also zero, we conclude that the right hand side of  Eq.\eqref{contourEQ} is  zero for $t<0$.

For $t>0$ we complete the contour by a counterclockwise tour at infinity in the upper half plane because it has zero contribution to the integral. Now the contour encloses both poles and  the Cauchy's residue theorem implies that
    \be
    \label{x-Fourier}
    x(t)=\frac{v_0}{2\omega_0}\oint\frac{d\Omega}{2\pi}e^{\ri\Omega t}\left(\frac{1}{\Omega+\omega_0}-\frac{1}{\Omega-\omega_0}\right)=\frac{\ri v_0}{2\omega_0}(e^{-\ri\omega_0t}-e^{\ri\omega_0t})=\frac{v_0}{\omega_0}\sin(\omega_0t).
    \ee
\begin{figure}[t]
  \centering
  \includegraphics[width=0.5\linewidth]{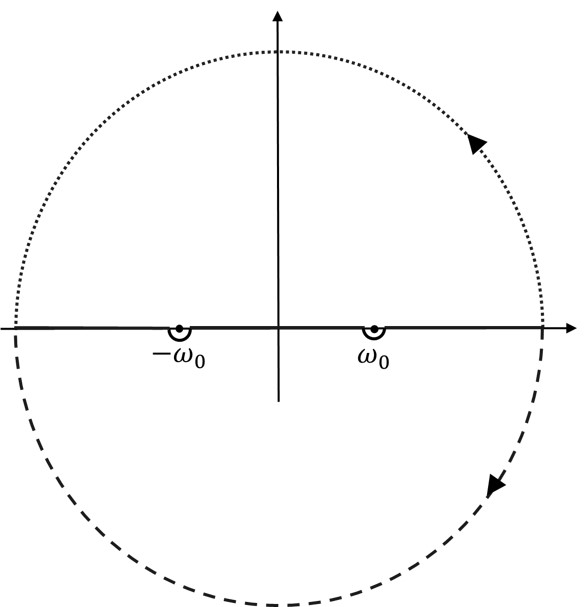}\\
  \caption{The useful contours to evaluate the integral \eqref{contourEQ}. The dotted and dashed
paths are used for $t>0$ and $t<0$ respectively.}\label{contourFig}
\end{figure}
So for $t>0$ we suppose that
    \be
    \label{delta-2}
    \frac{1}{\Omega^2-\omega_0^2}=2\ri\pi\delta(\Omega^2-\omega_0^2).
    \ee
We will use this recipe in the following.

What we have learned from this calculation is that the poles of $\tilde{x}(\Omega)$ correspond to the frequencies of oscillation of $x(t)$. Therefore, to obtain the  true frequencies of the system when an interaction  is added to the theory, we have to find the changes in the pole of the Green function. We begin by computing the Fourier transform of the equation
    \be
    \ddot x+\omega_0^2x+\lambda x^3-v_0\,\delta(t)=0.
    \ee
After a straightforward calculation, we obtain
    \be
    \label{main-Fourier}
    (\omega_0^2-\Omega^2)\,\tilde{x}(\Omega)+
    \lambda\int_{-\infty}^\infty\frac{d\Omega_1}{2\pi}\int_{-\infty}^\infty\frac{d\Omega_1}{2\pi}
    \tilde{x}(\Omega_1)\,\tilde{x}(\Omega_2)\,\tilde{x}(\Omega-\Omega_1-\Omega_2)-v_0=0.
    \ee
As this equation can not be solved exactly, we have to study it pertubatively. We do the perturbation in Fourier space, that is, we write
    \be
    \label{x-tilde}
    \tilde{x}(\Omega)=\tilde{x}^{(0)}(\Omega)+\lambda\tilde{x}^{(1)}(\Omega)+\cO(\lambda^2),
    \ee
in which, following \eqref{delta-1} and \eqref{delta-2},
    \be
    \label{x-tilde-0}
    \tilde{x}^{(0)}(\Omega)=\frac{v_0}{\omega_0^2-\Omega^2}=-\frac{\ri\pi v_0}{\omega_0}\left[\delta(\Omega-\omega_0)-\delta(\Omega+\omega_0)\right].
    \ee
Now we use \eqref{x-tilde}  in \eqref{main-Fourier} to obtain
    \be
    \label{x-tilde-1}
    (\omega_0^2-\Omega^2)\,\tilde{x}^{(1)}(\Omega)=-
    \lambda\int_{-\infty}^\infty\frac{d\Omega_1}{2\pi}\int_{-\infty}^\infty\frac{d\Omega_1}{2\pi}
    \tilde{x}^{(0)}(\Omega_1)\,\tilde{x}^{(0)}(\Omega_2)\,\tilde{x}^{(0)}(\Omega-\Omega_1-\Omega_2),
    \ee
where $\tilde{x}^{(0)}$ is given by \eqref{x-tilde-0}. Noting that the integrand on the RHS of \eqref{x-tilde-1} includes a sum over terms such as
    \be
    \delta(\Omega_1\pm\omega_0)\,\delta(\Omega_2\pm\omega_0)\,\delta(\Omega-\Omega_1-\Omega_2\pm\omega_0),
    \ee
we can compute it easily to obtain
    \bea
    (\omega_0^2-\Omega^2)\,\tilde{x}^{(1)}(\Omega)&=&
    \frac{\ri\pi\lambda v_0^3}{4\omega_0^3}\left[3\Big(\delta(\Omega-\omega_0)-\delta(\Omega+\omega_0)\Big)-
    \Big(\delta(\Omega-3\omega_0)-\delta(\Omega+3\omega_0)\Big)\right]\nn\\
    &=&    \frac{3\ri\pi\lambda v_0^3}{2\omega_0^2}\left[\delta(\Omega^2-\omega_0^2)-
    \delta(\Omega^2-9\omega_0^2)\right].
    \eea
Consequently
    \be
    \tilde{x}^{(1)}(\Omega)= \frac{3\ri\pi\lambda v_0^3}{2\omega_0^2}\left[\frac{\delta(\Omega^2-\omega_0^2)}{(\omega_0^2-\Omega^2)}+
    \frac{1}{8\omega_0^2}\delta(\Omega^2-9\omega_0^2)\right]=-\frac{3\lambda v_0^3}{4\omega_0^2}\left[\frac{1}{(\Omega^2-\omega_0^2)^2}+
    \frac{1}{8\omega_0^2}\frac{1}{9\omega_0^2-\Omega^2}\right],
    \ee
where we have used \eqref{delta-2} to obtain the second equality. Using this result together with  \eqref{x-tilde-0} in \eqref{x-tilde} we obtain
    \bea
    \tilde{x}(\Omega)&=&\frac{v_0}{\omega_0^2-\Omega^2}-\frac{3\lambda v_0^3}{4\omega_0^2}\left[\frac{1}{(\Omega^2-\omega_0^2)^2}+
    \frac{1}{8\omega_0^2}\frac{1}{9\omega_0^2-\Omega^2}\right]+\cO(\lambda^2)\nn\\
    &=&\frac{v_0}{\omega_0^2-\Omega^2+\frac{3\lambda v_0^2}{4\omega_0^2}}-
    \frac{3\lambda  v_0^3}{32\omega_0^4}\frac{1}{9\omega_0^2-\Omega^2}+\cO(\lambda^2).
    \eea
Thus, similarly to \eqref{x-Fourier}, the inverse Fourier transform \eqref{Fourier-inv} gives
    \be
    \label{x-Fourier-final}
    x(t)=\frac{v_0}{\omega}\sin\omega t-\frac{\lambda  v_0^3}{32\omega_0^5}\sin (3\omega_0t)+\cO(\lambda^2),
    \ee
where
    \be
    \label{omega-Fourier-final}
    \omega:=\sqrt{\omega_0^2+\frac{3\lambda v_0^2}{4\omega_0^2}}+\cO(\lambda^2)=\omega_0\left(1+\frac{3\lambda v_0^2}{8\omega_0^4}\right)+\cO(\lambda^2).
    \ee
It is interesting to see that \eqref{x-Fourier-final} and \eqref{omega-Fourier-final}  reproduce the results of section \ref{section-PL} for
    \begin{align}
    &A=\frac{v_0}{\omega},&\phi=-\frac{\pi}{2}.
    \end{align}

Here  the corrections to the frequency have been obtained through a  displacement in the location of the poles of the propagator
    \be
    \label{QFTrule}
    {\mbox{perturbation}}:\ \frac{e^{\ri\Omega t}}{\Omega^2-\omega_0^2}\ \ \ \rightarrow\ \ \ \frac{e^{\ri\Omega t}}{\Omega^2-\omega^2}.
    \ee

This approach resembles the  renormalization of the mass of particles in  QFT.
We encourage the reader to investigate the connection between the rule \eqref{QFTrule} and the rule of thumb introduced in subsection \ref{thumb}.
\subsection{Effective drive}

In this section, we would like to introduce another method to find the proper solution to the anharmonic oscillator perturbatively. Although  this method is not used in QFT, but we think it is worth mentioning. Again we begin with the fact that any solution to the equation of motion is periodic, although we do not know its period $T:=2\pi\omega^{-1}$. We only know that for $\lambda=0$ we have $T=2\pi\omega_0^{-1}$. In other words, we know that there exist a frequency $\omega$,  coefficients $a_n$ and phases $\phi_n$  such that
    \be
    \label{x-series}
    x(t)=a_1\cos(\omega t+\phi_1)+\lambda\sum_{n=2} a_n \cos(n\omega t+\phi_n),
    \ee
is the solution to the equation of motion. However, Eq.\eqref{x-series} can be interpreted as a solution to a driven SHO with frequency $\omega$ with an oscillatory drive $f(t)$,
    \be
    \label{x-drive}
    \ddot x+\omega^2x=f(t).
    \ee
To obtain a proper function $f(t)$ that reproduces the real solution of the equation of motion, we should insert \eqref{x-series} in \eqref{x-drive} which  gives
    \be
    f(t)=\lambda\sum_{n=2} (1-n^2\omega^2)\, a_n \cos(n\omega t+\phi_n).
    \ee
Since the higher frequency terms in \eqref{x-series} and, equivalently, in the drive $f(t)$  should be zero for $\lambda=0$, similarly to \eqref{delta-omega} we suppose that
    \begin{align}
    \label{a-phi}
    &a_n=a_n^{(0)}+\cO(\lambda),&\phi_n=\phi_n^{(0)}+\cO(\lambda),
    \end{align}
 with $a_1^{(0)}=A$ and $\phi_1^{(0)}=\phi$ according to \eqref{x0}. Equations \eqref{main} and \eqref{x-drive} give
    \be
    (\omega^2-\omega_0^2)x-\lambda x^3-\lambda\sum_{n=2} (1-n^2\omega^2) a_n \cos(n\omega t+\phi_n)=0.
    \ee
By using   \eqref{ansatz}, \eqref{x0},  \eqref{delta-omega} and \eqref{a-phi} we obtain
    \bea
    0&=&\mu A\cos(\omega_0t+\phi)-\frac{A^3}{4}\left\{\cos[3(\omega_0t+\phi)]+3\cos(\omega_0t+\phi)\right\}\nn\\&-&\sum_{n=2} (1-n^2\omega^2)\, a_n^{(0)} \cos(n\omega_0 t+\phi_n^{(0)})+\cO(\lambda),
    \eea
where we have also used \eqref{cos}. This gives
    \begin{align}
    &\mu=\frac{3A^3}{4},    &a_3^{(0)}=\frac{A^3}{32\omega^2},&&\phi_3^{(0)}=3\phi,
    \end{align}
and  $a_n^{(0)}=0$ for $n\neq 1,3$.
\subsection{Complex variables}
In this subsection, we combine some of the above ideas to devise a new  approach. This approach will help us in the next subsection in which we turn back to some QFT ideas.

Since the complex variables are the most natural representations of the Fourier coefficients, we rewrite Eq.\eqref{main} in terms of complex variables.  This can be done by considering two identical oscillators whose equation of motion are given by \eqref{main}:
    \begin{align}
    &\ddot x+\omega_0^2 x+\lambda x^3=0,& \ddot y+\omega_0^2 y+\lambda y^3=0.
    \end{align}
So $z:=x+{\rm i} y$ satisfies
    \begin{equation}
    \label{eom}
    \ddot z+\omega_0^2 z+\frac{3\lambda}{4}\left|z\right|^2z+\frac{\lambda}{4}{\bar{z}}^3=0.
    \end{equation}
This approach is quite useful. To obtain the first order corrections, we insert
    \begin{equation}
    z=Ae^{\ri\omega t}+\lambda f+{\cal{O}}(\lambda^2),
    \end{equation}
into Eq.(\ref{eom}) and obtain
    \begin{equation}
    \left(-\omega^2+\omega_0^2+\frac{3\lambda}{4}\left|A\right|^2\right)A\,e^{\ri\omega t}+\lambda\left(\ddot f+\omega_0^2f+\frac{{\bar{A}}^3}{4}\,e^{-3\ri\omega t}\right)+{\cal O}(\lambda^2)=0.
    \end{equation}
Therefore
    \begin{equation}
    \omega^2=\omega_0^2+\frac{3\lambda}{4}\left|A\right|^2,
    \end{equation}
and
    \begin{equation}
    f=\frac{{\bar{A}}^3}{4(9\omega^2-\omega_0^2)}e^{-3\ri\omega t}.
    \end{equation}
For $A\in\mathbb{R}$ we obtain
    \begin{equation}
    x=A\cos(\omega t)+\frac{\lambda A^3}{32\,\omega_0^2}\cos(3\omega t)+\cO(\lambda^2).
    \end{equation}
\subsection{Effective Action}
Now that we have rewritten the equation in terms of complex variables, we are ready to connect the problem to the finite temperature field theory,  in which one performs calculations with the same tools as in ordinary quantum field theory, but with compact Euclidean time \cite{Kaputsa}.

Let us examine the action of the system. Eq.(\ref{eom}) implies that the Lagrangian is
    \be
    L=\dot{z}\dot{\bar{z}}-\omega_0^2z\bar{z}-\frac{\lambda}{16}\left(z^4+\bar{z}^4+6z^2\bar{z}^2\right),
    \ee
hence the action is given by
    \be
    S=\int_{t_1}^{t_2}dt\,L.
    \ee
We know that the solution to Eq.\eqref{main} is oscillatory. Thus, we use the Fourier series in terms of a $\lambda$ dependent frequency. We insert this series into the Lagrangian $L$ and compute the mean value of the action in a period. This results in an action $\bar{S}$ for the Fourier coefficients whose minimum gives  the equations of motion in terms of the Fourier coefficients instead of $x(t)$.

We assume that the solution is an oscillatory one. Therefore, we have to minimize the action within the subspace of the functions which are periodic with frequency $\omega$. Therefore, it is suitable to rewrite the action as a function of the Fourier modes of $z$ and $\bar{z}$. Since we have restricted ourselves to the subspace of periodic functions, the action itself is also a periodic function of $t_1$ and $t_2$ and it is sufficient to consider the action within one period of motion. Therefore, we define the mean value of the action in a period
    \be
    \label{S-mean}
    \bar{S}:=\frac{\omega}{2\pi}\int_0^{\frac{2\pi}{\omega}}dt\,L,
    \ee
and rewrite it in terms of the Fourier coefficients of $z$ and $\bar{z}$ given by
    \begin{align}
    &z=\sum_{n\in\mathbb{Z}} z_ne^{\ri n\omega},
    &\bar{z}=\sum_{n\in\mathbb{Z}}\bar{z}_ne^{-\ri n\omega}.
     \end{align}
with $z_0=\bar{z}_0=0$. A straightforward calculation gives
    \bea
    \label{S-bar}
    \bar{S}&=&\sum_{n\in\mathbb{Z}}(n^2\omega^2-\omega_0^2)\,\omega_0^2z_n\bar{z}_n\nn\\&-&
    \frac{\lambda}{16}\sum_{\ell,m,n\in\mathbb{Z}}\left(z_\ell z_m z_n z_{-(\ell+m+n)}+
    \bar{z}_\ell \bar{z}_m \bar{z}_n \bar{z}_{-(\ell+m+n)}+6 z_\ell z_m \bar{z}_n\bar{z}_{\ell+m-n}\right),
    \eea
whose minimum gives the on-shell value of the Fourier modes. Since we are going to solve these equations by perturbation,  we  assume that
    \be
    \label{omega-alpha}
    \omega^2=\omega_0^2\left(1+\sum_{i=1}\lambda^i\mu^{(i)}\right),
    \ee
and
    \begin{align}
    \label{zn-series}
    &z_n=\sum_{i=0}\lambda^iz_n^{(i)},    &\bar{z}_n=\sum_{i=0}\lambda^i\bar{z}_n^{(i)}.
    \end{align}
For $\lambda=0$ the system reduces to an SHO with frequency $\omega_0$, we write  $z_n^{(0)}=\bar{z}_n^{(0)}=0$ for $n\neq 1$. We consider $z_1$ and $\bar{z}_1$ as benchmarks, so we take them to be  independent of $\lambda$, i.e., $z_1^{(i)}=\bar{z}_1^{(i)}=0$ for $i\neq 0$.
Therefore, $\bar{S}$ is a function of $z_1$, $\bar{z_1}$, and $z_n^{(i)}$, $\bar{z}_n^{(i)}$ for $n\neq1$, $i\ge1$.

Expanding the action  \eqref{S-bar} in powers of $\lambda$ we obtain
    \bea
    \bar{S}&=&\left(\lambda\mu^{(1)}+\lambda^2\mu^{(2)}\right)\omega_0^2z_1\bar{z}_1-\frac{3\lambda}{8}z_1^2\bar{z}_1^2
    +8\omega_0^2 \lambda^{2}z^{(1)}_{-3}\bar{z}^{(1)}_{-3}-
    \frac{\lambda^{2}}{4}\left(z_1^3z^{(1)}_{-3}+\bar{z}_1^3\bar{z}^{(1)}_{-3}\right)\nn\\&+&
    \lambda^2\sum_{n\not\in \{1,-3\}}(n^2-1)\,\omega_0^2z^{(1)}_n\bar{z}^{(1)}_n+\mathcal{O}(\lambda^3),
    \eea
and the equation of motion can be obtained order by order. It can  be verified that
    \be
    {z}^{(1)}_n=\bar{z}^{(1)}_n=0, \ \ \  n\not\in\{1,-3\}.
    \ee
while the cases $n=1$ and $n=3$ give
    \begin{align}
    &\mu^{(1)}=\frac{3}{4\omega_0^2}z_1\bar{z}_1,
    &\mu^{(2)}=\frac{3}{4\omega_0^2}\frac{\bar{z}_1^2\bar{z}^{(1)}_{-3}}{z_1},
    \end{align}
and
    \be
    z^{(1)}_{-3}=\frac{\bar{z}_1^3}{32\omega_0^2},
    \ee
implying that
    \be
    \mu^{(2)}=\frac{3}{128\omega_0^4}z_1^2\bar{z}_1^2.
    \ee

Interestingly we are obtaining the perturbative value of the frequency up to the second order of $\lambda$. To compute $\omega$ in terms of the non-perturbative amplitude $R$ we suppose $z_1=A$ and consequently
    \be
    R=A+\frac{\lambda A^3}{32\omega_0^2}+\mathcal{O}(\lambda^2),
    \ee
implying that
    \be
    A=R\left(1-\frac{\lambda R^2}{32\omega_0^2}\right)+\mathcal{O}(\lambda^2).
    \ee
Therefore, the correction to the frequency up to the second order of $\lambda$ is found to be
    \bea
    \frac{\omega^2}{\omega_0^2}&=&1+ \frac{3\lambda}{4\omega_0^2}R^2\left(1-\frac{\lambda R^2}{32\omega_0^2}\right)^2+\frac{3\lambda^2R^4}{128\omega_0^4}+\mathcal{O}(\lambda^3)\nn\\
    &=&1+ \frac{3\lambda}{4\omega_0^2}R^2-\frac{3\lambda^2R^4}{128\omega_0^4}+\mathcal{O}(\lambda^3),
    \eea
in agreement with \eqref{second-order}. The reader may consult \cite{GilmartinKleinandLi(1979)} for a similar approach.

\section{Coupled Oscillators}
Many systems can be modeled as a set of oscillators coupled to each other. The dependence of the frequency of each oscillator  on the amplitudes of the other oscillators is an interesting phenomenon. For example, consider $N$ pendulums attached to each other by a web of springs as depicted in Figure \ref{2pendulums} for $N=2$.  This system and many more similar examples are studied in the standard text books on classical mechanics such as \cite{Marion}. In general, the ``linearized'' equation of motion is
    \begin{align}
    &\sum_{j=1}^N \left(m_{ij}\ddot x_j+K_{ij}x_j\right)=0,& i=1,\cdots,N.
    \label{N-coupled-1}
    \end{align}
For simplicity, suppose  $m_{ij}=m\delta_{ij}$ for some $m>0$ where  $\delta_{ij}$ denotes the components of the identity matrix. Using the similarity transformation $U$, which makes the matrix of the coupling constants diagonal
    \begin{align}
    &K_{ij}\to k_i\delta_{ij}:=\sum_{m,n=1}^NU^{-1}_{im}K_{mn}U_{nj},
    &x_i\to c_{i0}:=\sum_{m=1}^NU_{im}x_m,
    \end{align}
and defining the bare frequencies $\nu_i:=\sqrt{k_i/m}$ we can rewrite the equation of motion \eqref{N-coupled-1} in the normal coordinates $c_i$
    \begin{align}
    &\ddot c_{i0}=-\nu_i^2 c_{i0},&i=1,\cdots N,
    \label{N-coupled-2}
    \end{align}
whose solution is
    \begin{align}
    \label{N-coupled-0-solution}
    &c_{i0}=A_i\cos(\nu_it+\phi_i),&i=1,\cdots,N.
    \end{align}

To study the effect of nonlinear contributions to the equation of motion, inspired by the setup depicted in Figure \ref{2pendulums}, we add third order terms in $x_i$ for $i=1.,\cdots,N$  to \eqref{N-coupled-1}. The similarity transformation then gives
    \begin{align}
    &\ddot c_i=-\nu_i^2 c_i-\lambda\sum_{jk\ell=1}^N V_{ijk\ell}\,c_jc_kc_\ell,&i=1,\cdots N,
    \label{N-coupled-3}
    \end{align}
in which $\lambda V_{ijk\ell}$ can be interpreted as the coupling constant which couples the normal modes $c_j$, $c_k$ and $c_\ell$ to each other. We consider $\lambda$ as the perturbation parameter. To solve \eqref{N-coupled-3}  perturbatively, we suppose
    \begin{align}
    &c_i=c_{i0}+\lambda\, c_{i1}+\cdots\,.
    \end{align}
Using this in \eqref{N-coupled-3} and keeping only  the zeroth order terms in $\lambda$ we obtain \eqref{N-coupled-2}. Thus the first order term in $\lambda$ reads
    \begin{align}
    &\ddot c_{i1}+\nu_i^2c_{i1}=-\sum_{jk\ell=1}^N V_{ijk\ell}\,c_{j0}c_{k0}c_{\ell0},&i=1,\cdots,N.
    \end{align}
Eq.\eqref{N-coupled-0-solution} implies that the ``drive force'' term on the RHS of this equation is a linear combination of oscillatory terms with frequencies
    \begin{align}
    \label{N-coupled-4}
    &\omega^{ijk}_{\alpha\beta\gamma}=\alpha\nu_{j}+\beta\nu_{k}+\gamma\nu_{\ell},&\alpha,\beta,\gamma=\pm1.
    \end{align}
If there is no degeneracy, i.e., if for all $i,j,k,\ell$ with $V_{ijk\ell}\neq0$ we have
    \be
    \omega^{ijk}_{\alpha\beta\gamma}\neq \nu_{i}.
    \ee
Then we can solve \eqref{N-coupled-4} easily. In the following, we study a few simple  degenerate systems.
\subsection{Degenerate  systems: Case I}
The simplest degenerate setup corresponds to $V_{iijj}\neq 0$ because in this case
    \be
    \omega^{ijj}_{++-}=\omega^{ijj}_{+-+}=\nu_{i}.
    \ee
\begin{figure}
\centering
  \includegraphics[width=0.4\linewidth]{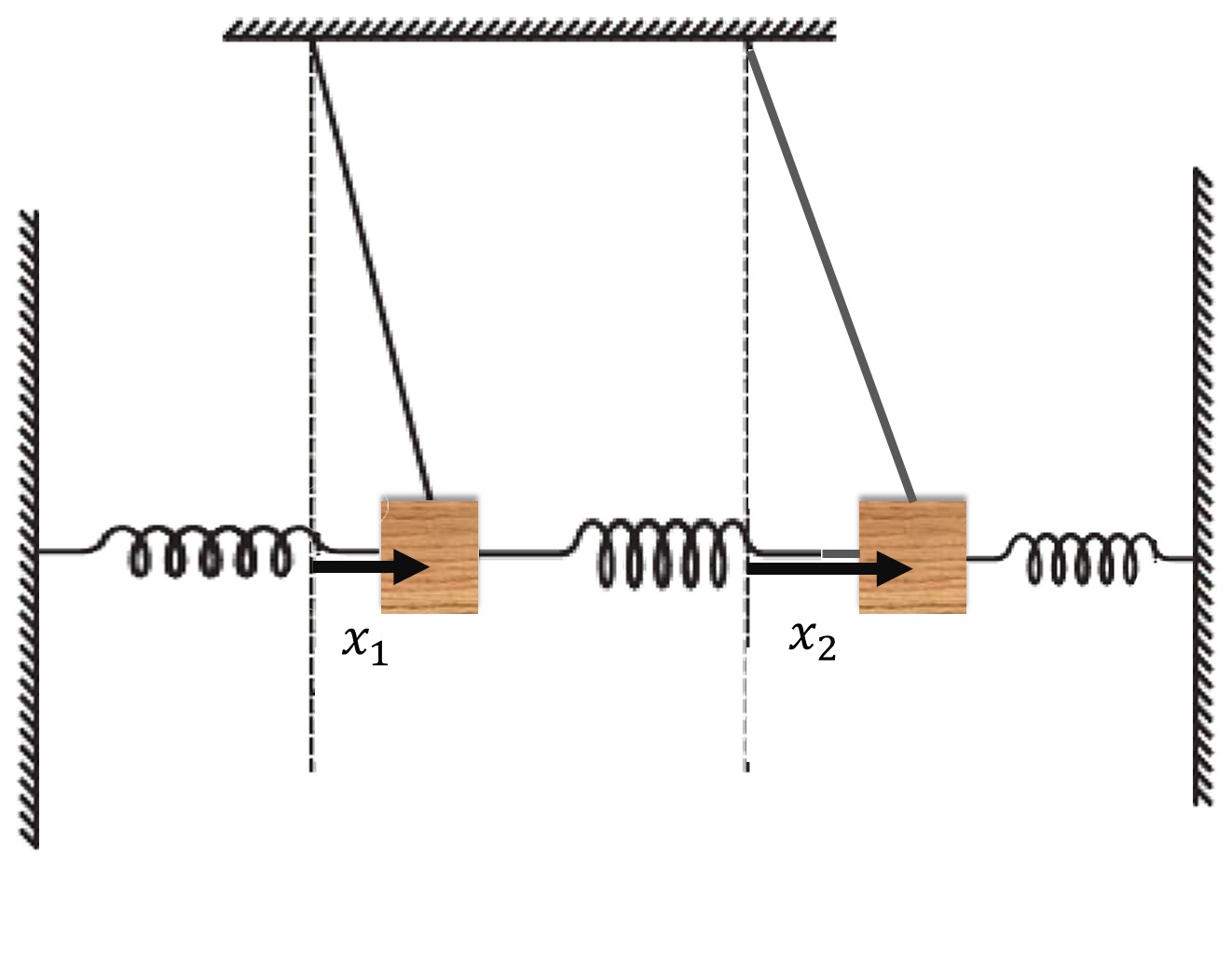}\\
  \caption{The system of two pendulums connected to  springs. The displacement
of the pendulums from their equilibrium  positions is shown with arrows.}\label{2pendulums}
\end{figure}

As an example, consider the system depicted in Figure \ref{2pendulums} with two identical pendulums with mass $m$ and length $L$ and three identical springs with stiffness $k$. If we denote the displacement of the pendulums from their equilibrium positions by $x_1$ and $x_2$ (indicated by horizontal black arrows in the figure) the normal coordinates would be
    \begin{align}
    &c_1:=x_1+x_2,&c_2:=x_1-x_2,
    \end{align}
and the equations of motion read
    \begin{align}
    \label{Degenerate1}
    &\ddot c_1+\nu_{1}^2c_1+\lambda\, c_1(c_1^2+3c_2^2)=0,
    &\ddot c_2+\nu_{2}^2c_2+\lambda\, c_2(3c_1^2+c_2^2)=0,
    \end{align}
where $\nu_{1}:=\sqrt{k/m+\mathfrak{g}/L}$, $\nu_{2}:=\sqrt{3k/m+\mathfrak{g}/L}$, and $\lambda:=\mathfrak{g}/24L^2$.

 Inspired by subsection \ref{thumb}, we suppose that the  true frequency of the leading term of $c_1$ and $c_2$ is $\omega_1$ and $\omega_2$ respectively which are different from the corresponding bare values by an amount of order $\lambda$. Therefore, we suppose  that
    \begin{align}
    &c_i=A_i\cos(\omega_it+\phi_i)+\mathcal{O}(\lambda),&i=1,2.
    \end{align}
 Using this in \eqref{Degenerate1} we obtain
    \begin{align}
    &\omega_{1}^2=\nu_{1}^2+\frac{3\lambda}{4}\left(A_1^2+2A_2^2\right),
    &\omega_{2}^2=\nu_{2}^2+\frac{3\lambda}{4}\left(2A_1^2+A_2^2\right).
    \end{align}

Therefore, the change in the frequency of a mode may come not only from its own amplitude, but also the amplitudes of other modes may have a contribution to it. This kind of degeneracy is widespread in usual systems, for example, one may think of systems of $N>2$ pendulums connected to one another by springs.  However, there is another kind of degeneracy which leads to a plethora of leading order oscillations which we discuss  in the following.

\subsection{Degenerate  systems: Case II}
We encounter a rich structure of ``leading order frequencies'' in systems which in addition to the $V_{iijj}$ couplings have other sources of degeneracy. For example, consider the Lagrangian
    \begin{align}
    &L=\sum_{n=1}^4\left(\frac{1}{2}{\dot x}_n^2-\frac{\nu_n^2}{2}x_n^2-\frac{\lambda}{4} x_n^4\right)-\lambda V_4,&V_4:= x_1x_2x_3x_4,
    \end{align}
where $\nu_n:=n\omega_0$, $n=1,2,3,4$. The equations of motion are
    \begin{align}
    &\ddot x_n+\nu_n^2 x_n+\lambda x_n^3+{\lambda}\frac{x_1x_2x_3x_4}{x_n}=0.
    \end{align}
 Obviously this system is degenerate because $\nu_{2}+\nu_{3}-\nu_{4}=\nu_{1}$. In addition, because of the existence of $x_n^3$ term, the degeneracy discussed in the previous section is present in the system.
Assume that to the leading order,  different modes of the system oscillate with frequencies $\omega_n=n\omega_0+\lambda\varpi_n$. The degeneracy coming from the $\lambda x_n^3$ term can be removed by tuning $\varpi_n$. Since this tuning depends on the amplitude of the leading order terms, we anticipate that for generic initial conditions there should be no further degeneracy from the  $V_4$ interaction.

Although this argument is correct, the contribution from the $V_4$ interaction generates more leading order terms. To see this, focus on the equation of motion for $x_1$. The $V_4$ interaction gives a term whose frequency is
    \begin{align}
    &\tilde{\omega}_1:=\omega_2+\omega_3-\omega_4=\omega_1+\lambda\tilde\varpi_1,&\tilde\varpi_1:=\varpi_2+\varpi_3-\varpi_4-\varpi_1.
    \end{align}
This means that we should add to $x_1$ the solution of the  equation
    \be
    \ddot{\tilde x}_1+\omega_1^2\tilde x_1=-\lambda\frac{A_2A_3A_4}{4}\cos(\tilde\omega_1t+\phi_2+\phi_3-\phi_4),
    \ee
which is
    \begin{align}
    &\tilde x_1=\tilde{A}_1\cos(\tilde\omega_1t+\phi_2+\phi_3-\phi_4),&\tilde{A}_1:=-\frac{\lambda A_2A_3A_4}{4(\omega_1^2-{\tilde\omega}_1^2)}=\frac{ A_2A_3A_4}{8\omega_1\tilde\varpi_1}+\mathcal{O}(\lambda).
    \end{align}
So the denominator $(\omega_1^2-{\tilde\omega}_1^2)$ is proportional to $\lambda$  and the envisaged first order correction   actually becomes a zeroth order term. In other words, by following the usual procedure we have generated a new leading order term instead of a sub-leading term. This term oscillates with a frequency close to $x_1$ and results in a  beating phenomenon. Starting with \(\tilde{\omega}_1\) and repeating the calculation,  i.e., changing $\tilde\omega_1$ to remove the degeneracy from the $x_1^3$ term, a new leading order term may emerge. Therefore, this procedure is successful only if there are finitely many  leading order terms. Otherwise, we should devise a new approach to the problem.

A simple example in which the above-mentioned procedure fails definitely is  a system with three degrees of freedom whose equations of motion are
    \begin{align}
    \label{b1}
    &\ddot x_i+\nu^2_i x_i=-\lambda\prod_{j\neq i}x_j,&i=1,2,3,
    \end{align}
where $\nu_2=\nu_1+\nu_3$.  To solve these equations, following  subsection \ref{thumb}, we consider the ansatz
    \be
    \label{b2}
    x_i=\sum_{m=1}^NA_{im}\cos(\omega_{im} t+\phi_{in})+\lambda\sum_{m=1}^{N'}B_{im}\cos(\Omega_{im} t+\theta_{in}),
    \ee
in which
    \begin{align}
    & \lim_{\lambda\to 0}\omega_{in}= \nu_i,&\lim_{\lambda\to 0}\Omega_{in}\neq \nu_i.
    \end{align}
So naively we are anticipating at most $N$ leading order terms corresponding to each degree of freedom.
\eqref{b2} solves \eqref{b1} only if for any $m$ there  exist $n_q$ and $p_q$ such that
    \be
    \omega_{2m}=\omega_{1n_q}+\omega_{3p_q},\ \ \ q\in\{1,\cdots N\}.
    \ee
This implies that for  generic values of the initial conditions  $N$ is not finite.

\begin{figure}[t]
  \centering
  \includegraphics[width=0.5\linewidth]{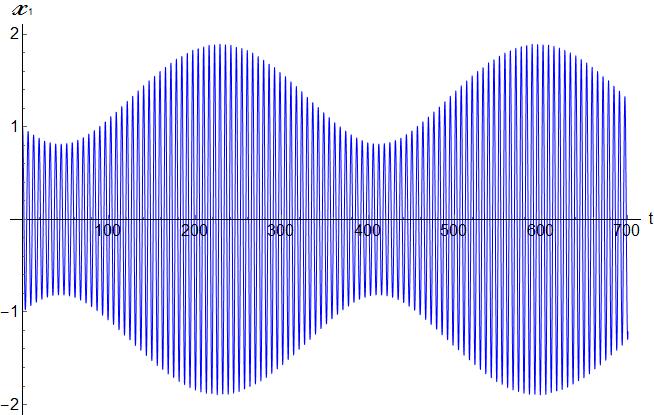}\\
  \caption{Numerical solution of \eqref{b1} for $\nu_1=\nu_2/3=\nu_3/2=1$ and $\lambda=-
0.03$. The initial conditions are $x_1(0)=1$, $\dot x_1(0)=0$, $x_2(0)=0$, $\dot x_2(0)=3$,
and $x_3(0)=1$, $\dot x_3(0)=1$. The beating pattern indicates the existence of at least two
leading order terms with slightly different frequencies and amplitudes.}\label{Beat}
\end{figure}

\begin{figure}[t]
\centering
  \centering
  \includegraphics[width=0.5\linewidth]{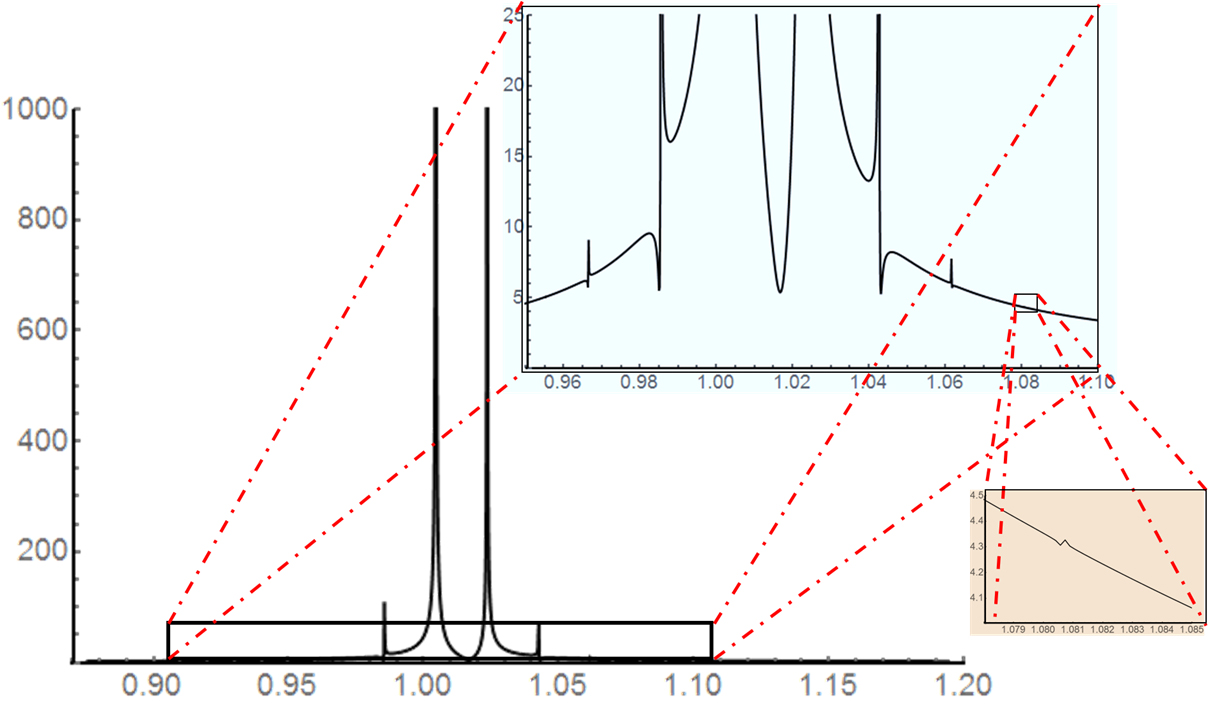}\\
  \caption{The numerical Fourier transform of the solution depicted in Figure \ref{Beat}. In the main plot, four different peaks are observed. In the inset
graphs, in which we have zoomed in the original graph, four more frequencies
can be identified, all within a few percent away from the bare frequency (=1). }
  \label{FourierFIG}
\end{figure}

Figure \ref{Beat} shows a numerical solution of \eqref{b1} for $t\in[0,700]$. The beat phenomenon observed in the solution confirms the existence of two or more oscillations with close frequencies. The beat frequency is completely sensitive to  the initial conditions and the parameter $\lambda$, which supports the idea we have developed.  The corresponding Fourier transform is depicted in Figure \ref{FourierFIG} for 40000 temporal steps. The unit of the vertical axis is arbitrary. We recognize four frequencies on the large-scale diagram. More frequencies can be detected by focusing. Therefore, the zeroth order of the solution can have several oscillations with frequencies very close the original one.

\section{Summary}

Learning quantum field theory, especially the renormalization procedure, is usually difficult for the first time. Students usually do not have the insight into why counter-terms are added to the theory and hardly understand how such terms remove the singularities. Sometimes they wonder if we are really following a justifiable procedure, that is, when we are following a standard perturbative expansion, why some singular terms appear. More surprisingly, why these singularities disappear when some physical quantities are ``renormalized''. In this paper, we have applied a similar  procedure to an anharmonic oscillator, a  simple and familiar system in classical mechanics. We have arrived at some singular terms while performing a perturbative appoach. However, due to simplicity of the system, one can easily identify the source of such singularities. Then we have explained why and how adding a counter-term  to the theory resolves the problem. The crucial point is to distinguish between a parameter of the theory ($\omega_0$ in the linear system) and a physical quantity (the period of the oscillations of the non-linear system). Similarly, in QFT we  distinguish between the mass parameter of the free theory and the physical quantity (mass) in the interacting theory which takes correction when interaction is added.
  
Moreover, in order to make  the similarities with quantum field theory  manifest, we have devised some other methods, in addition to the well-known methods, in line with quantum field theory. These new methods include the use of different tools and concepts in QFT such as Green function, the propagators and effective action. 

Finally, we have considered a system of coupled oscilators to see how nonlinear terms affect their physical quantities. In addition to similar findings in the single oscillator case, we have found an interesting phenomenon in degenerate coupled oscillatory systems: the oscillations of the system will be composed of a number of frequencies close to each frequency of the linear system. This leads to an oscillatory motion with beats for each normal mode of the system.  We have  justified our perturbative result numerically. 


\end{document}